\documentclass[fleqn,12pt,twoside]{article}
\usepackage{espcrc1}

\usepackage{graphicx}
\usepackage[figuresright]{rotating}

\newcommand{\AmS}{{\protect\the\textfont2
  A\kern-.1667em\lower.5ex\hbox{M}\kern-.125emS}}

\title{$\Delta G$ from high $p_T$ events at SMC and high $p_T$ analysis at COMPASS}
\author{K.Kurek\address[sins]{Andrzej So{\l}tan Institute for Nuclear Studies,
\\ Ho\.za 69, 00-681 Warsaw, Poland.}
\thanks{ Talk given at 10$^th$ International Conference Baryons 04, October 25-29, 2004,
 Ecole Polytechnique, Palaiseau, France. Supported in part by the by SPUB NO.
621/E-78/SPB/CERN/P-03/DWM 576/2003-2006}\\ On behalf of SMC and
COMPASS Collaborations}

\begin{document}

\maketitle

\begin{abstract}

Measurements of the longitudinal spin cross section asymmetry for
deep inelastic muon-nucleon interactions with two high transverse
momentum hadrons ($p_T >$ 0.7 GeV) in the final state are
presented for SMC data for polarized proton and deuteron and for
data on polarized deuteron from COMPASS taken in 2002 and 2003.
The muon asymmetries determined with a cut on $Q^2>$ 1 GeV$^2$ in SMC are:
$A_p = 0.03 \pm 0.057\pm 0.01$ and $A_d = 0.070 \pm 0.076 \pm
0.010$, respectively. From these values a gluon polarization
$\Delta G /G = -0.20\pm 0.28\pm 0.10$ was obtained at an average
fraction of nucleon momentum carried by gluons $\eta = 0.07$.  The
measured asymmetry (with cut on $Q^2>$ 1 GeV$^2$) in COMPASS 
is $(A_d/D) = -0.015 \pm 0.08 \pm
0.013$ where D is depolarization factor and the gluon polarization
$\Delta G /G = 0.06\pm 0.31\pm 0.06$  was obtained at $\eta =
0.13\pm 0.08$.
\end{abstract}

\section{Introduction.}
The measurement of the asymmetry for the pairs of hadrons with the
high transverse momenta is a way of direct measurement of the
gluon polarization via the Photon Gluon Fusion (PGF) process~\cite{carlitz,pt1}.
Requiring two hadrons with high transverse momenta in the final state the PGF process
contribution, originally small in DIS, can be increased.

\begin{figure}
\begin{center}
\includegraphics[width=10cm]{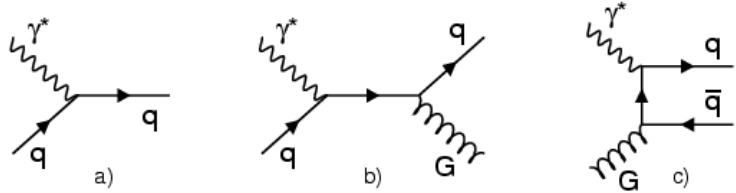}
\end{center}
\caption{Lowest order diagrams for DIS $\gamma^{\ast}$ absorption:
a) leading process (LP),
b) gluon radiation (QCDC),
c) photon-gluon fusion (PGF). \label{fig:pgf}}
\end{figure}

The cross section asymmetry with the production of at least two
hadrons with large transverse momenta,  $A^{\ell N \rightarrow
\ell hhX}$, at the parton level consists of three terms
corresponding to the contributions from three lowest order
processes: leading process (LP), PGF and QCD Compton process
(QCDC):
\begin{equation}
 A^{\ell N \rightarrow \ell hhX} =
  {\frac{\Delta q} {q}}( \langle\hat{a}_{LL} \rangle^{LP}  R_{LP} +
  \langle\hat{a}_{LL}\rangle^{QCDC}  R_{QCDC} ) +
{\frac{\Delta G} {G}}\langle\hat{a}_{LL}\rangle^{PGF} R_{PGF}\;.
\label{eq:gluon}
\end{equation}
The relevant diagrams are schematically shown in Fig.\ref{fig:pgf}.

In the asymmetry formula $\langle\hat{a}_{LL}\rangle$ is the
average partonic asymmetry for a given process  and   $R$ the
ratio of its cross section with respect to the total cross section
for the selected sample of events. The measurement of the
asymmetry  $A^{\ell N \rightarrow \ell hhX}$  thus permits an
evaluation of the gluon polarization, $\Delta G/G$,  if all other
elements in Eq.~(\ref{eq:gluon}) are known. The quark asymmetry
$\Delta q / q$ can be approximated from the value of $A_1$
obtained in inclusive measurements. The partonic asymmetries
$\hat{a}_{LL}$ and the ratios $R$ are calculated for Monte Carlo
simulated events in the kinematic region covered by the data. The
applicability of the presented method is based on the assumption
that three considered processes can be separated and therefore it
is limited to the leading order QCD approximation. The description
of hadron production in DIS muon interaction, in terms of the
three processes showed in Fig.~\ref{fig:pgf} using Monte Carlo
event generator LEPTO has been successfully tested in EMC and E665
experiments. Other processes, such as those involving resolved
photon contribution, are expected to have  small contributions for
$Q^2$ above $1$~GeV$^2$ and have not been taken into account in
the presented analysis.

The cross section asymmetry, $A^{\ell N \rightarrow \ell hhX}$, is
related to the counting rate  asymmetry, $A^{exp}$,  by:
$A^{exp}=P_B P_T f A^{\ell N \rightarrow \ell hhX}$, where $f$ is
the effective dilution factor, which takes into account the
dilution of spin asymmetries by unpolarized nuclei in the target
and by radiative effects on the nucleon, $P_B$ and $P_T$ stand for
beam and target polarization, respectively.

\section{Data selection. }
Event sample for further studies was defined by requiring 2
hadrons with transverse momenta above 0.7~GeV and energy fraction
$z$ above 0.1 (to reduce target fragmentation contribution). Also
selection on event kinematics, including $Q^2\ge 1$~GeV$^2$ and
vertex position were applied.

The SMC experiment has used polarized muon beam of CERN SPS with
energy 190 GeV scattered off polarized proton (butanol or amonia)
and deuteron (deuterated butanol) targets. The high intensity
polarized muon beam with energy 160 GeV is used in COMPASS
experiment together with polarized $^6$LiD target. The $^6$LiD
target with higher dilution factor $f$ (larger fraction of the
polarized deuterons inside the target) and high muon intensity
beam allow to increase the precision of the asymmetry measurement
in COMPASS.

In the SMC experiment after all selections the total number of
remaining events amounts to about 80k for the proton and  70k for
the deuteron sample. The detailed discussion of the SMC analysis
is presented in~\cite{pt_paper}.

The data after selection were compared with the MC sample
generated with LEPTO and the  experimental conditions have been
taken into account. In order to describe the data, it was found
necessary to change the values of two  parameters describing the
quark fragmentation in JETSET~\cite{pt_paper}. The statistical
precision of the  gluon polarization determined from
Eq.~(\ref{eq:gluon}) depends on the precision of the  measured
asymmetry $A^{\ell N \rightarrow \ell hhX}$ and on the fraction of
PGF events ($R_{PGF}$) in the final sample. Several methods of
additional selections were tried to obtain a finale sample large
enough and with a maximal contribution of PGF events.

\begin{figure}
\begin{tabular}{p{0.47\textwidth}p{0.47\textwidth}}
\includegraphics[width=7.1cm]{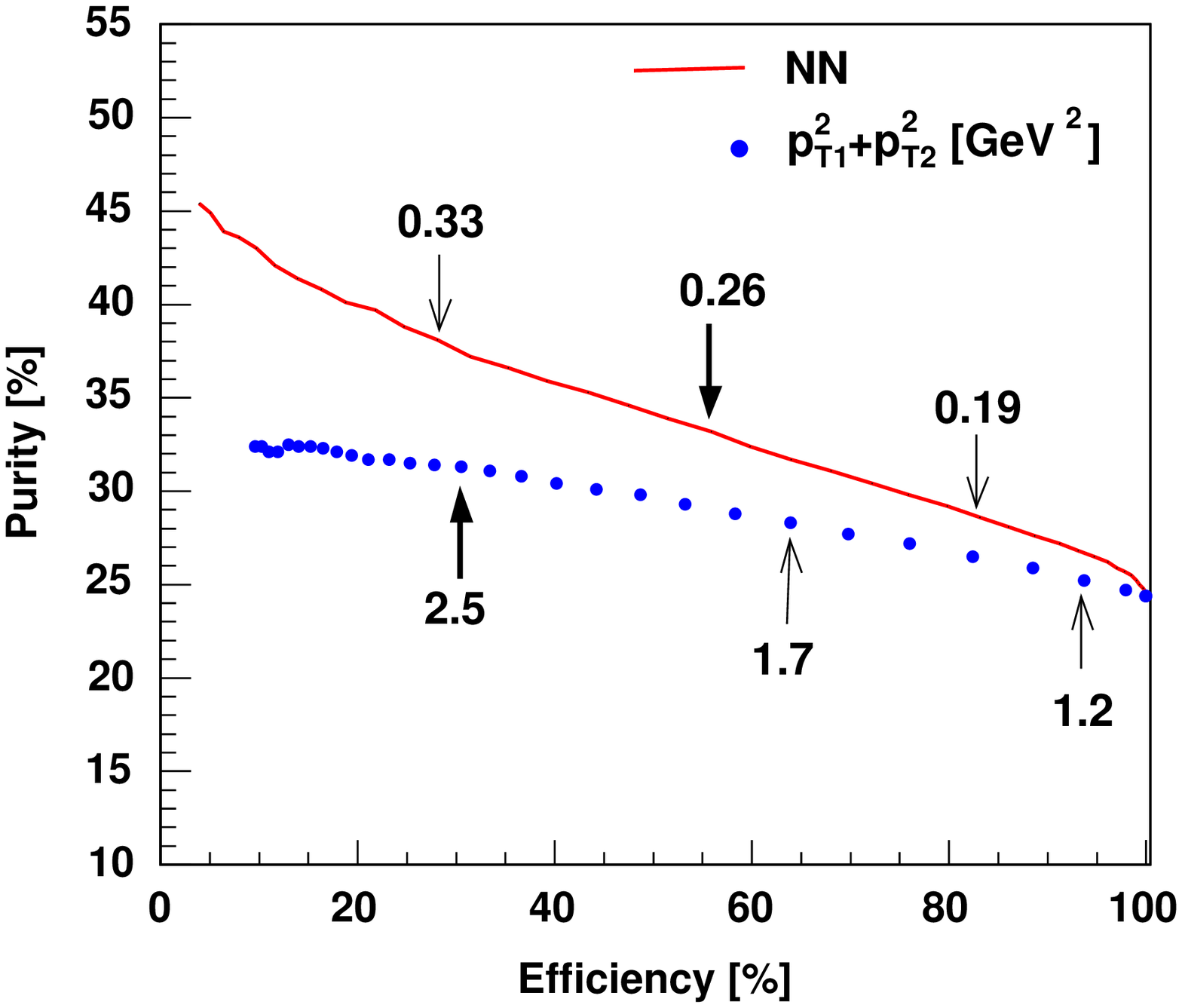}&
\includegraphics[width=6.5cm]{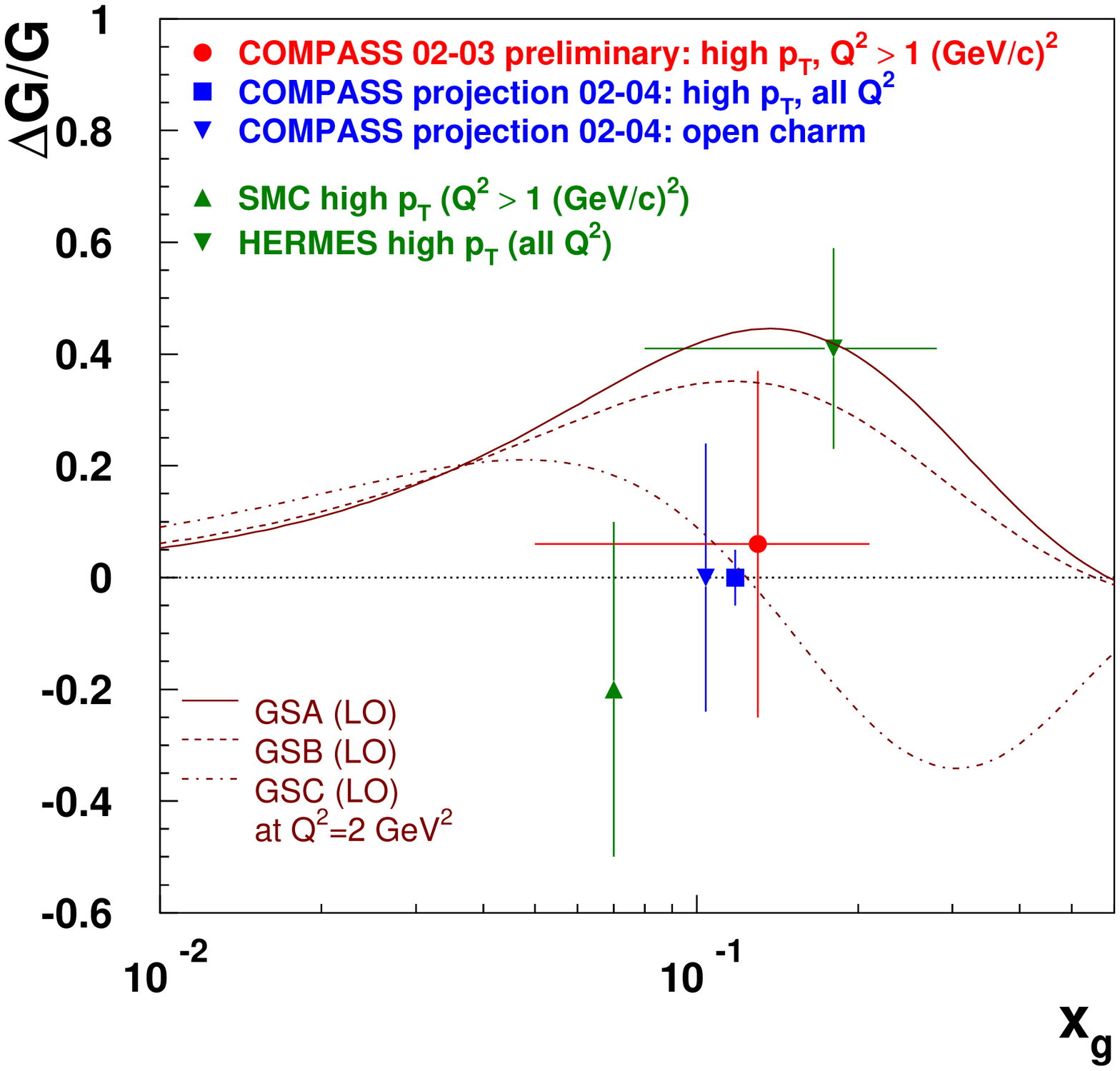}
\vskip-1pc \\ \caption{Purity and  efficiency obtained in the
selection of the PGF process with the methods based on the cut on
$\sum p_T^2$ and the NN response, applied to simulated events for
SMC experiment.} \label{fig:comp_traf} & \caption{SMC and COMPASS
results for $\Delta G/G$  from high $p_T$ hadron pairs with
$Q^2>1$ GeV$^2$ compared with Hermes result (all $Q^2$). In
additon projected errors for open charm and high $p_T$ hadron
pairs for all $Q^2$ from COMPASS  2002-2004 data are shown. }
\label{fig:dg}\\
\end{tabular}
\vskip-1pc
\end{figure}

For SMC data the best results were obtained with the selection
based on the neural network (NN) approach ~\cite{acta_paper}.
Comparison of selection efficiency and sample purity for this
method and a method using cut on the sum $p_{T1}^{2}+p_{T2}^{2}$
is shown in Fig.\ref{fig:comp_traf}. Better  statistical precision
was obtained with the neural network method. For COMPASS the
selection $p_{T1}^{2}+p_{T2}^{2} > 2.5$ GeV$^2$ was used.

\section{Gluon polarization determination from SMC data. }

The gluon polarization has been evaluated from Eq.~(\ref{eq:gluon}) using
the measured  $A^{\ell N \rightarrow \ell hhX}$ asymmetry, obtained for
the samples with enhanced $R_{PGF}$ by NN selection.
 The measured muon asymmetries are: $A_p = 0.03 \pm
0.057\pm 0.01$ and $A_d = 0.070 \pm 0.076 \pm 0.010$. The
partonic asymmetries  $\hat{a}_{LL}$ of each  sub-process have
been calculated for each Monte-Carlo event  and averaged. The
values of the ratios $R$ provided by the simulation for the LP,
QCDC and PGF processes are 0.38, 0.29 and 0.33, respectively. The
contributions  of different processes for the proton and deuteron
samples differ by less than 0.02.

The  gluon polarization is determined for the  kinematic region
covered  by the selected sample and corresponds to a given
fraction of nucleon momentum carried by gluons $\eta$. This
quantity is known for simulated events  but cannot be directly
determined from the data. The averaged value for the selected PGF
events in the MC sample, $\eta =$0.07, is used as the reference
value for the result obtained on $\Delta G/G$.

Averaging the results for  proton and deuteron  obtained with the
neural network selection  ${{\Delta G} /  {G}} = -0.20 \pm 0.28
\pm 0.10$ has been obtained. The accuracy is limited by the
reduction to less than 1\% of the DIS sample by the hadron
selection requirements. The systematic errors contain dependence
on the simulation parameters, scale dependence in the generation
and systematic uncertainty on the measured asymmetry.

\section{The high $p_T$ analysis at COMPASS and the result for gluon polarization.}

The asymmetry for the selected sample with $p_{T1}^{2}+p_{T2}^{2}
> $2.5 GeV$^2$ was calculated with the additional cut $x_Bj <0.05$.
This cut selects kinematical region where asymmetry from LP and
QCDC are small (proportional to $A_1^d$) and these processes
contribute only as a dilution to the measured signal.

To remove the contribution from resonances an invariant mass of two
hadrons were required to be higher than $1.5$ GeV.
The obtained asymmetry  for events with $Q^2>$ 1 GeV$^2$ is $A_d/D = -0.015 \pm 0.08 \pm 0.013$, where D is a depolarization factor.
The systematic uncertainty takes into account false asymmetries, the uncertanties  from the measurements of the
target and beam polarizations.

To determine $\Delta G/G$ from the measured asymmetry the LEPTO
generator has been used with the SMC modifications in JETSET quark
fragmentation functions and with radiative corrections included
(RADGEN). The PGF fraction $R_{PGF}=
\frac{\sigma^{PGF}}{\sigma^{tot}}$, has been found to be $0.34\pm
0.07$.

The obtained gluon polarization is ${{\Delta G} /  {G}} = 0.06 \pm
0.31 \pm 0.06$ for  $\eta =0.13\pm 0.08$(rms).

The high $p_T$ sample without  $Q^2$  cut (all $Q^2$) contains roughly 10 times more data than the sample with $Q^2 >$ 1 GeV$^2$.
The preliminary result for asymmetry (2002 data only) gives the value $(A_d/D) = -0.065 \pm 0.036\pm 0.01$.

\section{Summary.}
 In the Fig.~\ref{fig:dg} SMC, COMPASS and
Hermes ~\cite{hermes} results on ${{\Delta G} /  {G}}$ are
presented together with the projected errors from open charm and
high $p_T$ hadron pairs for all $Q^2$ from COMPASS  2002-2004
data. The results for the gluon polarization ${{\Delta G} /  {G}}$
from  SMC and COMPASS analysis (using 2002-2003 data) with $Q^2>$
1 GeV$^2$ are consistent with zero. Based on error projections,
use of all $Q^2$ sample will allow a more precise determination of
the gluon polarization on COMPASS in the near future.

\end{document}